\documentclass[aps, pra, twocolumn, floatfix, superscriptaddress, longbibliography]{revtex4-1}
\usepackage{amsmath}
\usepackage{amsfonts}
\usepackage{amssymb}
\usepackage{hyperref}
\hypersetup{colorlinks=true,
allcolors=blue}
\usepackage{eucal}
\usepackage{xspace}

\usepackage{graphicx}

\newcommand{\dttw}{\ensuremath{\Delta E(\omega,\tau)/E(\omega)}}
\newcommand{\scoc}{Sr$_2$CuO$_2$Cl$_2$\xspace}
\newcommand{\lco}{La$_2$CuO$_4$\xspace}
\newcommand{\nco}{Nd$_2$CuO$_4$\xspace}
\newcommand{\ybco}{YBa$_2$Cu$_3$O$_6$\xspace}
\newcommand{\cuo}{CuO$_2$\xspace}

\begin{document}

\title{Transient terahertz photoconductivity of insulating cuprates}

\author{J.~C.~Petersen}
\affiliation{Department of Physics, Simon Fraser University, Burnaby, BC, Canada}
\author{A.~Farahani}
\affiliation{Department of Physics, Simon Fraser University, Burnaby, BC, Canada}
\author{D.~G.~Sahota}
\affiliation{Department of Physics, Simon Fraser University, Burnaby, BC, Canada}
\author{Ruixing Liang}
\affiliation{Department of Physics and Astronomy, University of British Columbia, Vancouver, BC, Canada}
\affiliation{Canadian Institute for Advanced Research, Toronto, ON, Canada}
\author{J.~S.~Dodge}
\email{jsdodge@sfu.ca}
\affiliation{Department of Physics, Simon Fraser University, Burnaby, BC, Canada}
\affiliation{Canadian Institute for Advanced Research, Toronto, ON, Canada}

\begin{abstract}
We establish a detailed phenomenology of photocarrier transport in the copper oxide plane by studying the transient terahertz photoconductivity of \scoc  and \ybco. The peak photoconductivity saturates with fluence, decays on multiple picosecond timescales, and evolves into a state characterized by activated transport. The time dependence shows little change with fluence, indicating that the decay is governed by first-order recombination kinetics. We find that most photocarriers make a negligible contribution to the dc photoconductivity, and we estimate the intrinsic photocarrier mobility to be 0.6--0.7~$\text{cm}^2/\text{V}\,\text{s}$ at early times, comparable to the mobility in chemically doped materials.
\end{abstract}


\maketitle

\section{Introduction}
\label{sec:intro}
How does a charge carrier move in an antiferromagnetic insulator? Since Brinkman and Rice posed this deceptively simple question over forty years ago~\cite{Brinkman1970}, it has become a canonical problem in condensed matter physics and is widely thought to be at the heart of high-temperature superconductivity~\cite{Lee2006}. Attempts to answer it have traditionally focused on chemically doped systems with one type of carrier. But it is also possible to examine the motion of photoexcited charge carriers, which can now be studied experimentally in real time with pulsed lasers~\cite{Ulbricht2011, Basov2011, Wall2012, Giannetti2016}, and theoretically with advanced computational methods~\cite{Eckstein2013, Iyoda2014, Kanamori2011, Aoki2014}.
As a step in this effort, we have measured the transient terahertz photoconductivity (THz-PC) of insulating antiferromagnetic copper oxides, and characterized its dependence on time, frequency, excitation density, and temperature. We find that photocarriers evolve from a nonequilibrium state of relatively high mobility to a thermalized state that has an activated temperature dependence. These measurements provide a benchmark for future research on optical and single-particle excitations in magnetic solids.

The photoinduced absorption (PIA) spectrum of copper oxides at $\hbar\omega \geq 0.1$~eV resembles the absorption spectrum of chemically doped compounds~\cite{Kim1991, Matsuda1994, Perkins1998, Okamoto2010, Okamoto2011}, and suggests the existence of a short-lived Drude conductivity peak~\cite{Okamoto2011}. But at lower frequencies, photoconductivity measurements have been limited to nanosecond timescales and are dominated by impurity-band conduction~\cite{Thio1990}. With its picosecond time resolution, THz-PC provides a way to probe the intrinsic low-frequency charge dynamics of photocarriers in the copper oxide plane~\cite{Ulbricht2011, Basov2011}. By eliminating the influence of chemical doping, it complements transport and optical measurements on lightly doped cuprates~\cite{Ando2001, Sun2004, Mishchenko2008}. It also supplements photoemission measurements, which provide access to single-particle states but not to their transport properties~\cite{Damascelli2003, Armitage2010}.

We focus here on photoexcitations in cuprates, but our work connects to broader research on nonequilibrium phenomena in strongly correlated systems~\cite{Okamoto2010, Okamoto2011, Iwai2003, Okamoto2007, Wall2011, Giannetti2016, Aoki2014}. In some cases, for example, photocarrier injection can induce transitions between distinct structural, magnetic, conducting, and superconducting phases~\cite{Okamoto2011, Okamoto2010, Yonemitsu2008}. Photoexcitation can also liberate bound carriers by melting an ordered state that inhibits their conduction in equilibrium~\cite{Basov2011}. Finally, a growing body of work focuses on similar experiments with quantum gases in optical lattices~\cite{Esslinger2010}.

\section{Photoexcitations in insulating copper oxides}
\label{sec:photocarriers}

Before we present our results, we describe the basic features of photoexcitations in the \cuo plane and discuss their evolution as the system returns to equilibrium. To relate this to photoconductivity, we adopt a semiclassical description in which photocarriers exhibit Boltzmann kinetics and a quasi-equilibrium Drude-Lorentz response to the terahertz field~\cite{Ulbricht2011}. The carriers are regarded as either free or bound, with population densities determined by the initial photoexcitation and subsequent kinetic processes. An important assumption of this picture is that the Drude-Lorentz relaxation times are short compared to the pump-probe time delay~\cite{Orenstein2015}, which is justified empirically by our results.

\begin{figure}
	\centering
	\includegraphics[width=\columnwidth]{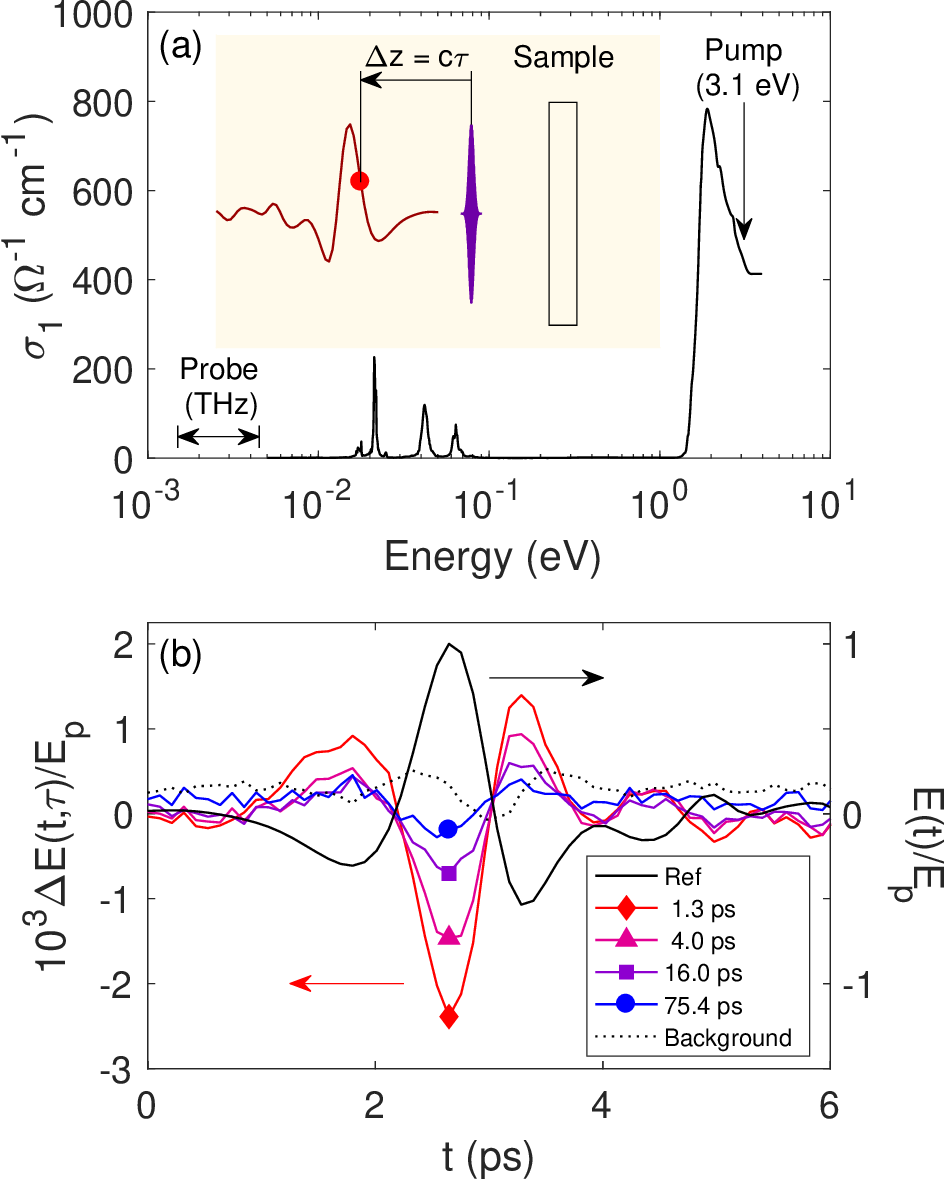}
	\caption{(color online). Transient terahertz photoconductivity of Sr$_2$CuO$_2$Cl$_2$. (a)~Equilibrium optical conductivity from Zibold et al.~\cite{Zibold1996}, annotated to indicate the visible pump photon energy and the terahertz probe photon energy range. Inset:~Schematic of the experimental geometry. The pump excites photocarriers in the sample, and the terahertz probe pulse measures the photoconductivity as a function of time. After passing through the sample, the terahertz waveform $E(t)$ is measured in the time domain by repetitive sampling of the electric field at the terahertz measurement time $t$; the red marker indicates the point associated with $t = 2.3$~ps. We measure the change in $E(t)$ in response to a pump pulse that precedes the measured field point by a time delay $\tau$, to determine $\Delta E(t,\tau)$. (b)~Measurements at pump fluence $F = 245~\mu\text{J/cm}^2$ of $\Delta E(t,\tau)$ relative to the background (obtained at $\tau = -14.7$~ps) for $\tau =$~1.3, 4.0, 16.0, and 75.4~ps. Also shown is the reference field $E(t)$, which is transmitted through the sample without pump excitation. All curves are normalized to $E_p$, the peak value of $E(t)$, with different vertical scales for $\Delta E(t,\tau)/E_p$ (left) and $E(t)/E_p$ (right), as indicated by the arrows.}
	\label{fig:time}
\end{figure}

\begin{figure}
	\centering
	\includegraphics[width=\columnwidth]{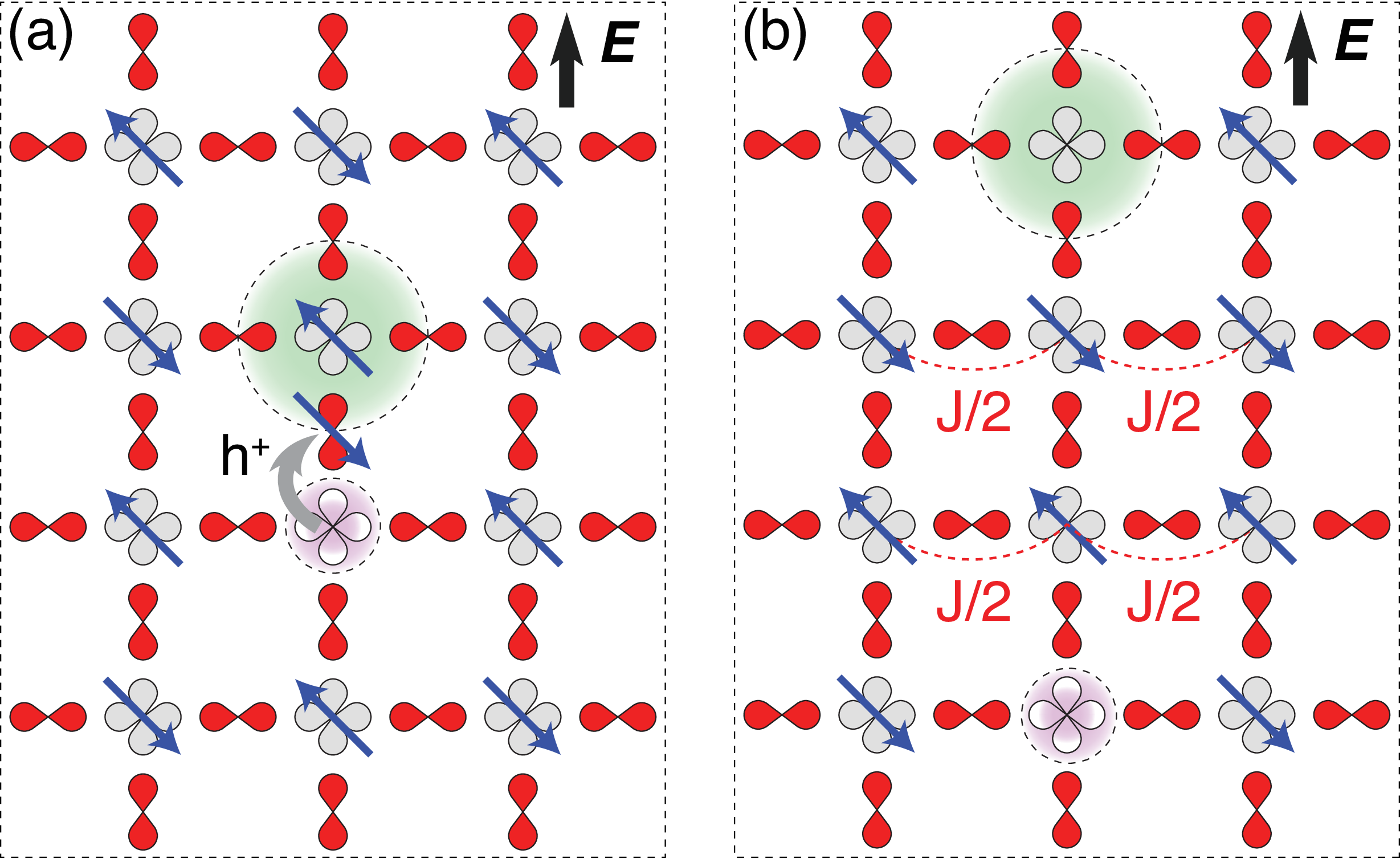}
	\caption{(color online). Schematic of the photoexcitation process in the \cuo plane, simplified by ignoring lattice distortions and quantum fluctuations of both spin and charge. Singly occupied Cu($3d_{x^2-y^2}$) orbitals are gray, and $\sigma$-bonded O($2p$) orbitals are red. Blue arrows indicate the spin orientation of holes that occupy each Cu($3d_{x^2-y^2}$) orbital in an antiferromagnetic arrangement at equilibrium. (a)~The pump field transfers a hole from a Cu($3d_{x^2-y^2}$) orbital to a neighboring O($2p$) orbital. The pink circle denotes the negatively-charged spinless copper site, while the green circle indicates positive charge distributed among symmetry-equivalent oxygen orbitals. The copper and oxygen holes form a Zhang-Rice spin singlet~\cite{Zhang1988}, represented as spinless in the subsequent panel. (b)~A probe electric field, applied in the direction of the black arrow in the upper right corner of each panel, causes the photoexcited charge carriers to move, which in turn disrupts the antiferromagnetic order. Dotted red lines indicate unfavorable magnetic bonds that each cost an energy $J/2$.}
	\label{fig:schem}
\end{figure}

Undoped \cuo compounds are antiferromagnetic insulators with an optical gap near 2~eV~\cite{Zaanen1985, Tokura1990, Cooper1993, Zibold1996, Kastner1998}. Figure~\ref{fig:time}(a) shows the optical conductivity of a prototypical copper-oxide insulator, \scoc~\cite{Zibold1996}, and Fig.~\ref{fig:schem} shows a simplified picture of the photoexcitation process. Optical excitation above the gap transfers an electron from a O($2p$) orbital to a nearby Cu($3d_{x^2-y^2}$) orbital to form a spinless Cu($3d^{10}$) state, leaving the oxygen hole to form a spinless Zhang-Rice singlet by binding with an adjacent copper hole~\cite{Zhang1988}. Antiferromagnetic spin excitations couple strongly to the motion of both charge excitations, as illustrated in Fig.~\ref{fig:schem}(b). We refer to these electron-like and hole-like excitations as ``doublons'' and ``holons,'' respectively, to emphasize their composite character and distinguish them from their independent-electron-theory counterparts.

Mutual Coulomb attraction and magnetic interactions may also bind photocarriers into excitons~\cite{Simon1996}, and both optical measurements and resonant inelastic x-ray spectroscopy~(RIXS) show a broad peak near the charge-transfer gap that resembles one~\cite{Tokura1990, Falck1992, Kastner1998, Ellis2008, Ament2011}. The RIXS peak also shows a direct-gap excitation with a minimum energy at the $\Gamma$ point~\cite{Ellis2008}, whereas a simple convolution of the single-particle holon and doublon dispersion relations would produce an indirect gap~\cite{Armitage2010, Damascelli2003}. Yet the same measurements indicate that any bound state at this energy is short-lived, if it exists at all. The RIXS peak loses spectral weight as its momentum increases away from $\Gamma = 0$, indicating that it merges with the single-particle continuum over much of the Brillouin zone~\cite{Ellis2008}. And over the entire Brillouin zone, the two-particle linewidth is comparable to and overlaps with the single-particle angle-resolved photoemission spectroscopy (ARPES) dispersion, suggesting that the two-particle states are better described as short-lived resonances than as true bound states~\cite{Falck1992, Ellis2008, Armitage2010, Damascelli2003}.

All of this indicates that excitons will form a negligible fraction of the initial photoexcited-state population, especially if we photoexcite well above the gap energy, as we do in our experiments. If we define the quantum yield $\Phi$ to be the fraction of photoexcitations that initially form mobile holon-doublon pairs, this means $\Phi\approx 1$. The photoconductivity is given by $\sigma = n_{c} e\bar{\mu} = 2\Phi n_\text{ex} e\bar{\mu}$, where $n_c = n_h + n_d$ is the total photocarrier density, $\bar{\mu} = (\mu_h + \mu_d)/2$ is the average mobility of free holons and doublons, and $n_\text{ex}$ is the initial photoexcitation density. In our experiments we photoexcite below the threshold photon energy for impact ionization~\cite{Werner2014}, so we also expect $\Phi\leq 1$. Assuming $\Phi \lesssim 1$, a measurement of $\sigma$ places bounds on the larger of the two carrier mobilities, $2\bar{\mu}\ge \max(\mu_h,\mu_d)\ge \bar{\mu}$.

As Fig.~\ref{fig:schem} suggests, holons and doublons have dramatically different spatial wavefunctions, so we might expect $\mu_h$ and $\mu_d$ to reflect that difference. But a variety of theoretical calculations yield similar quasiparticle dispersion relations for the two carrier types~\cite{Lee2006}, and ARPES measurements on both electron-doped and hole-doped cuprates support this conclusion~\cite{Armitage2010, Damascelli2003}. Holons and doublons appear to couple to phonons with comparable strength~\cite{Mishchenko2010}, and their transport mobilities both fall in the range $\mu\approx 1\text{--}10~\text{cm}^2/\text{V\,s}$~\cite{Ando2001, Sun2004}. With this rationale, we may take $\mu_d \approx \mu_h \approx \bar{\mu}$ and $\Phi\approx 1$ to estimate the mobility directly from the initial photoconductivity.

Intraband relaxation and interband recombination return the system to equilibrium over multiple timescales. Model calculations indicate that photocarriers relax their excess energy to the spin and lattice degrees of freedom during the first few hundred femtoseconds after photoexcitation~\cite{Yonemitsu2009, DeFilippis2012, Eckstein2013, Kogoj2014, Golevz2014, Eckstein2016}. Mid-infrared PIA measurements show a subpicosecond initial decay that has been attributed to nonradiative recombination through multimagnon emission~\cite{Okamoto2010, Okamoto2011, Lenarvcivc2013, Lenarvcivc2014a}. The signal subsequently decays nonexponentially over several hundred picoseconds; one interpretation of this is that carriers that do not recombine in the first few hundred femtoseconds become less mobile, creating a bottleneck to further recombination by reducing the holon-doublon encounter rate~\cite{Okamoto2010, Okamoto2011}.  Conventional photoconductivity measurements show that some of these residual photocarriers can remain for several seconds~\cite{Thio1990}.

The initial PIA decay rate is independent of excitation density~\cite{Okamoto2010, Okamoto2011}, which implies first-order recombination kinetics. Extended-state recombination is at least second order in the carrier density, so the initial recombination must proceed through some kind of localized state. This should be reflected in the photoconductivity. For example, if holon-doublon pairs form excitons before recombining~\cite{Lenarvcivc2013, Lenarvcivc2014a}, we expect the photoconductivity to be suppressed at frequencies below the exciton binding energy before the recombination event depletes the PIA at higher frequencies~\cite{Kaindl2003, Kaindl2009}. Alternatively, if individual holons or doublons enter traps before recombining with oppositely charged carriers in extended states~\cite{Shockley1952,Hall1952}, we would expect a Drude peak from the extended carriers and Lorentzians from the traps, and they would decay together. Finally, holon-doublon pairs may remain weakly bound by their mutual Coulomb attraction while retaining the ability to respond to a terahertz probe pulse as quasifree carriers~\cite{Knoesel2004}. Here, we would expect a broad Drude-like peak to decay through diffusion-limited recombination.

Experimentally, it is difficult to distinguish between exciton-mediated, trap-mediated, or diffusion-limited recombination in the cuprates. The PIA spectrum below 0.5~eV decays more rapidly than elsewhere, but this is dominated by much larger changes in the overall mid-IR PIA spectral weight during the first few picoseconds~\cite{Okamoto2010, Okamoto2011}. In conventional semiconductors, spectral weight can shift between a free-carrier Drude peak and a Lorentzian-like exciton peak over a timescale that is much shorter than the recombination time, making it possible to identify the spectral change with a shift in occupation~\cite{Kaindl2003, Kaindl2009}. In cuprates, by constrast, the redistribution is secondary to the changes in the overall spectrum. As a result, while the overall PIA spectrum shows evidence for recombination, and the spectral weight transfer indicates that photocarriers become localized over time, it is less clear whether these localized states are responsible for the first-order kinetics or merely a residue of it. 

To summarize, photoconductivity in insulating cuprates involves composite holon and doublon charge carriers which should have approximately equal mobilities. To the extent that excitonic effects are important, their binding energy is small enough that photocarriers excited well above the gap energy should initially be delocalized. During the same subpicosecond timescale that the carriers relax through intraband scattering, experiments show that the holons and doublons recombine through a first-order kinetic process that involves localized states; the remaining carriers then evolve over multiple timescales. Transient terahertz photoconductivity measurements can monitor this entire evolution in real time, and determine the photocarrier mobility at early times.

\section{Experiment}
\subsection{Sample preparation}
\label{sec:sample}
We studied single-crystal platelets of two prototypical insulating cuprates, \scoc (SCOC) and \ybco (YBCO6). All samples measured several millimeters in each $a$-$b$-planar dimension and a few tenths of a millimeter along the $c$ axis. Both YBCO6 and SCOC are tetragonal, so they are optically isotropic in the $a$-$b$ plane---unlike YBa$_2$Cu$_3$O$_{7-\delta}$, for example, which has one-dimensional Cu-O chains, is monoclinic, and is optically anisotropic~\cite{Zhao2016}. SCOC crystals were grown by cooling the melt from $1110~^{\circ}$C to $1075^{\circ}$~C at a rate of $3~^{\circ}\text{C}/\text{hr}$ in an alumina crucible~\cite{Muller-Buschbaum1977}. YBCO6 crystals were fabricated by a top-seeded melt growth technique~\cite{Liang2006}. The oxygen content of 6.00 to 6.01 in the YBCO6 crystal is achieved by annealing at $700~^{\circ}$C under $5 \times 10^{-7}$ torr oxygen pressure for 10 days. We saw no significant changes in the optical response during weeks of continuous experiments on the same samples, indicating the absence of photoinduced structural changes.

\subsection{Transient terahertz photoconductivity}
\label{sec:THz-PC}
The inset to Fig.~\ref{fig:time}(a) shows a schematic of our measurement process. We create photocarriers by illuminating the sample at a 10-degree angle of incidence with 100~fs, 3.10~eV pump pulses from a frequency-doubled laser amplifier that operates at a 1~kHz repetition rate. We probe their transient THz response using standard nonlinear optical methods with ZnTe generator and detector crystals~\cite{Ulbricht2011}. The optical penetration depth of the pump is approximately 100~nm, and the samples are transparent at terahertz frequencies, so we use a transmission geometry and treat the photoconducting layer in the thin-film approximation for analysis. The terahertz beam diameter is 2.2~mm or less for all frequencies above 2~meV, and our pump beam diameter is 5~mm. The noncollinear geometry limits our temporal resolution to approximately one picosecond.

To determine the time-dependent photoconductivity, we mechanically chop the pump beam to measure $\Delta E_\text{chop}(t,\tau)$, the change in the THz probe field as a function of THz measurement time $t$ and pump delay $\tau$. We observe a residual signal at negative pump delays, indicating that the system has not fully recovered during the 1~ms interval between successive THz pulses. We treat this as a background and subtract it from all measurements at positive delays, to obtain $\Delta E(t,\tau) = \Delta E_\text{chop}(t,\tau>0) - \Delta E_\text{back}(t)$, where $\Delta E_\text{back}(t) = \Delta E_\text{chop}(t,\tau=-14.7~\text{ps})$. We then compare this to a reference THz field $E(t)$ transmitted through the unexcited sample.

Figure~\ref{fig:time}(b) shows $E(t)$ together with $\Delta E(t,\tau)$ at various values of $\tau$, all for SCOC and normalized to $E_p$, the peak of the reference field at $t=t_p$. Results for YBCO6 are qualitatively similar, with quantitative differences that we discuss below. Dividing the difference signal by the reference in the frequency domain gives \dttw, the fractional change in the THz transmission spectrum. The optical penetration depth is much shorter than the THz penetration depth, so we treat the photoexcited material as an optically thin conducting film on an insulating substrate. In the small signal limit, and for time-delays greater than the photocarrier momentum relaxation time~\cite{Orenstein2015}, the THz-PC is $\sigma(\omega,\tau)\approx-(1+n)[\Delta E(\omega,\tau)/E(\omega)]/\delta Z_0$, where $Z_0$ is the impedance of free space, $n$ the THz refractive index of the sample, and $\delta$ its optical penetration depth; we took values for $n$ and $\delta$ from Refs.~\onlinecite{Zibold1996, Tajima1991, Kelly1989}.

\begin{figure}
	\centering
	\includegraphics[width=\columnwidth]{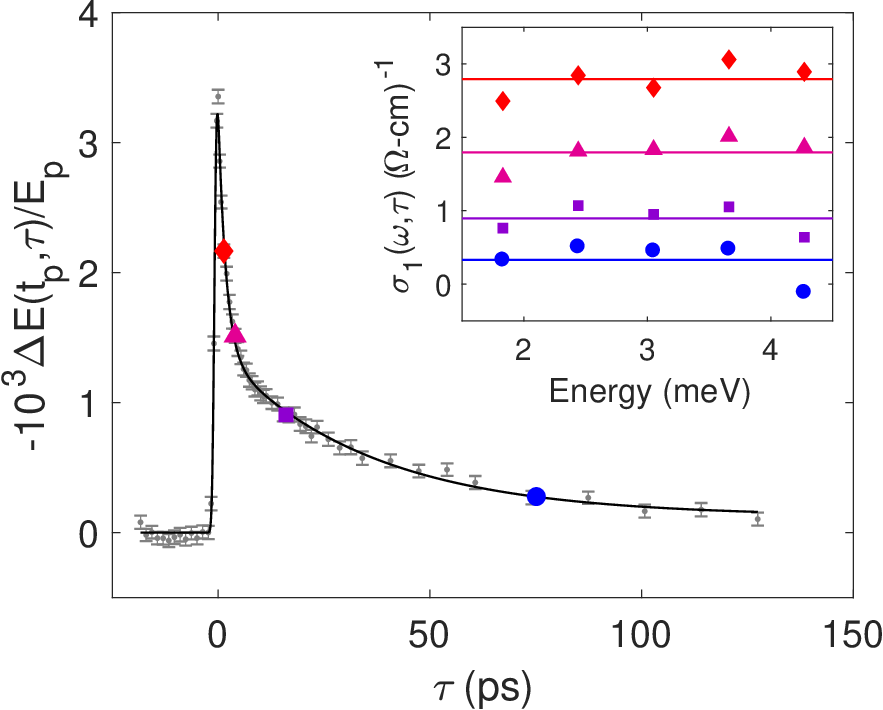}
	\caption{(color online). Room-temperature fractional change in the peak THz field amplitude transmitted through SCOC after photoexcitation with $F = 245~\mu\text{J/cm}^2$. Each point represents an average over $N>300$ individual measurements, and the error bars show the standard error, estimated with a bootstrap method. The solid line depicts a model of multiple exponential decays convolved with a Gaussian pulse shape, as described in the text. Markers identify $\tau = 1.3$~($\circ$), 4.0~($\diamond$), 16.0~($\triangledown$), and 75.4~($\Box$) ps, as in Fig.~\ref{fig:time}(b).  Inset:~Room temperature photoconductivity spectra obtained from Fig.~\ref{fig:time}(b). Solid lines indicate mean values.}
	\label{fig:decay}
\end{figure}

The inset of Fig.~\ref{fig:decay} shows the real part of the THz-PC, $\sigma_1(\omega,\tau)$, obtained from the data in Fig.~\ref{fig:time}(b). Our results for the imaginary part, $\sigma_2(\omega,\tau)$, are consistent with zero within our measurement uncertainty. This supports our assumption that the photocarrier momentum relaxation time is short, providing a consistency check on our analysis. We also find that $\sigma_1(\omega,\tau)$ is approximately constant across our measurement bandwidth, and remains so as it decays with increasing $\tau$. The spectra clearly extrapolate to a nonzero $\sigma_1$ at $\omega = 0$, indicating the existence of free, mobile carriers over the timescale of our measurement.

The lack of dispersion in $\sigma_1(\omega,\tau)$ allows us to focus on the time dependence by relating the THz-PC to the difference signal at the peak of the THz field. We define $\sigma_{1}(\tau) =  -(1+n)\Delta E(t_p,\tau)/\delta Z_0 E_p\approx \langle\sigma_1(\omega,\tau)\rangle_\omega$,  and show $-\Delta E(t_p,\tau)/E_p$ in the main panel of Fig.~\ref{fig:decay}. The THz-PC develops promptly upon photoexcitation, then decays on multiple picosecond timescales. As discussed in Sec.~\ref{sec:photocarriers}, qualitatively similar time dependence is observed in the mid-infrared PIA spectrum of other insulating cuprates~\cite{Okamoto2010, Okamoto2011}, indicating it is influenced mainly by the carrier density, not the mobility.

We can fit the time dependence with a model of multiple exponential decays convolved with a 1~ps FWHM Gaussian pulse shape, $y(\tau) = [f*g](\tau)$, with $f(t) = e^{-(t-t_0)^2/2w^2}/(\sqrt{2\pi}w)$, $w=425$~fs, and $g(t) = \Theta(t)(a_1e^{-t/T_1}+a_2e^{-t/T_2} + a_3)$. For the data shown in Fig.~\ref{fig:decay}, this yields time constants $T_1 = 1.8\pm 0.1$~ps and $T_2 = 36\pm 3$~ps (here and elsewhere uncertainties are given as standard errors), with $\chi^2 = 53$ for 57 time points and six fit parameters. Measurements on YBCO6 show similar behavior, with $T_1 = 1.1\pm 0.1$~ps, $T_2 = 7.7\pm 0.6$~ps, and $\chi^2 = 9$ for 42 time points and six fit parameters. Note that the measured conductivity is reduced by a factor of $\zeta = y(0)/g(0)$ from its intrinsic $w\rightarrow 0$ value at $\tau=0$ in this model; the fits give $\zeta = 0.73\pm 0.02$ for SCOC and $\zeta = 0.6\pm0.1$ for YBCO6.

\section{Results and discussion}
\label{sec:results}
\subsection{Peak photocarrier mobility}
\label{sec:peakmu}
\begin{figure}
	\centering
	\includegraphics[width=\columnwidth]{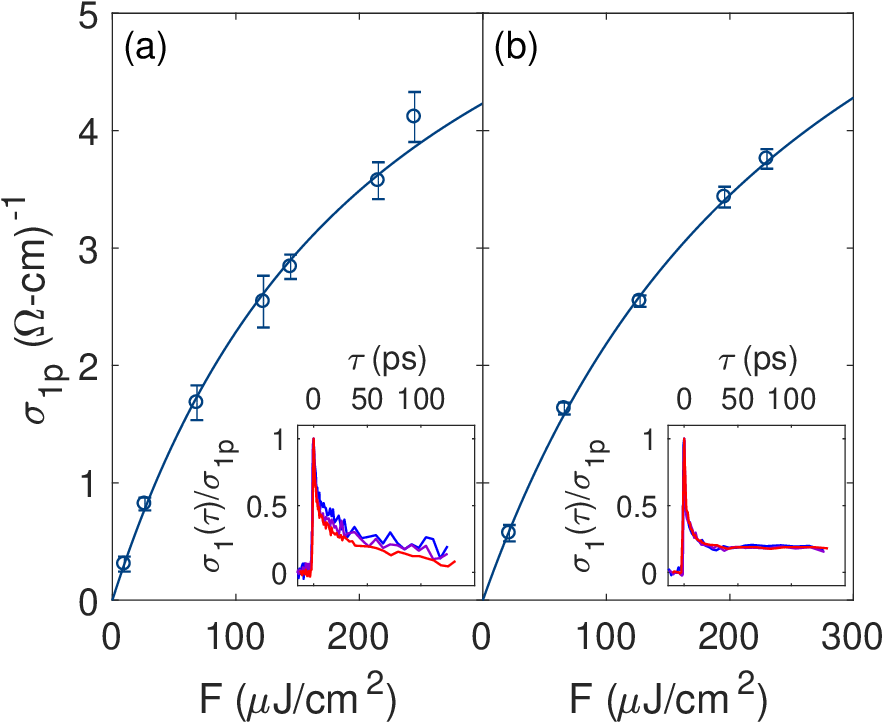}
	\caption{(color online). Peak photoconductivity as a function of incident fluence for (a) SCOC and (b) YBCO6. Solid lines show fits with the saturation model discussed in the text, with $F_0 = 221~\mu\text{J/cm}^2$ for SCOC and $F_0 = 275~\mu\text{J/cm}^2$ for YBCO6. The insets show the time-dependent photoconductivity, normalized to the peak value, for three different fluences: 69, 122, and 245 $\mu\text{J/cm}^2$ for SCOC and 66, 127, and 230 $\mu\text{J/cm}^2$ for YBCO6, denoted by blue, purple, and red curves, respectively. In YBCO6, the temporal response exhibits no significant fluence dependence, whereas in SCOC there is a weak increase in the overall decay rate with increasing fluence.}
\label{fig:fluence}
\end{figure}

We can estimate the photocarrier mobility by examining the peak THz-PC, $\sigma_{1p} = \sigma_{1}(\tau=0)$, as a function of incident fluence $F$, shown in Fig.~\ref{fig:fluence}. Measurements on both materials are fit well by a saturation model,
\begin{equation}
\sigma_{1p}(F) = \frac{\alpha F}{1 + F/F_0},
\label{eq:saturation}
\end{equation}
where $F_0$ is the saturation fluence and $\alpha$ is constant. Similar fluence dependence is visible in the pump-probe response of several correlated insulators at higher probe frequencies~\cite{Iwai2003, Okamoto2007, Uemura2008, Okamoto2011, Novelli2014}, and we believe it is caused by a new  recombination channel that opens at high fluence~\cite{Sahota}. At low fluence, $F\ll F_0$, $\sigma_{1p} \approx \alpha F$, and we can determine the mobility from
\begin{equation}
\alpha = \frac{\zeta\Phi(1-\mathcal{R})}{\delta\hbar\omega_p}e(\mu_d+\mu_h),
\label{eq:tpclimit}
\end{equation}
where $\zeta$ is the reduction factor defined in Sec.~\ref{sec:THz-PC}; $\Phi$ is the quantum yield defined in Sec.~\ref{sec:photocarriers}; $(1-\mathcal{R})/(\delta\hbar\omega_p) = n_\text{ex}/F$, the ratio of the excitation density to the fluence in the $F/F_0\rightarrow 0$ limit; $\mu_{d,h}$ are the doublon and holon mobilities; $\omega_p$ is the pump photon energy; and $\mathcal{R}$ is the equilibrium reflectance. Taking the values given in Sec.~\ref{sec:THz-PC} for $\zeta$ and assuming $\Phi=1$ and $\mu_d\approx\mu_h$, fits with Eqs.~(\ref{eq:saturation}) and~(\ref{eq:tpclimit}) yield $\mu_h = 0.72\pm 0.06~\text{cm}^2/\text{V\,s}$ for SCOC and $\mu_h = 0.6\pm 0.1~\text{cm}^2/\text{V\,s}$ for YBCO6. These estimates provide lower bounds on the intrinsic photocarrier mobility of the copper oxide plane, and they are six orders of magnitude larger than earlier photocarrier mobility estimates~\cite{Thio1990}. If $\mu_d\neq\mu_h$, the larger of the two mobilities could be as much as a factor of 2 greater, and if $\Phi<1$, the mobility will be $1/\Phi$ greater.

Our analysis assumes that the macroscopic photocarrier spatial distribution remains approximately constant during our measurement time, and our mobility measurements justify this assumption. Using the Einstein relation, we can estimate the carrier diffusion constant at room temperature to be $D \approx 1.9~\text{nm}^2/\text{ps}$, which implies that photocarriers travel $17~\text{nm}$ within the $a$-$b$ plane during our 150~ps observation window. Furthermore, dc resistivity measurements of lightly doped YBa$_2$Cu$_3$O$_{6.2}$ indicate that the mobility along the $c$ axis is close to one thousand times smaller than in the plane~\cite{Semba2001}. This reduces the out-of-plane diffusion distance to 0.5~nm, much smaller than the optical penetration depth of the pump.

We compare our results with the PIA measurements of \textcite{Okamoto2010,Okamoto2011} in Table~\ref{tab:compare}, although this involves significant systematic uncertainties. The most important of these is that we can not simply extrapolate Eq.~(\ref{eq:saturation}) to the higher densities used in their measurements, because we have found separately that the saturation fluence $F_0$ varies significantly with both material and pump wavelength~\cite{Sahota}, and both of these vary between the PIA measurements and ours. We extrapolate instead with $\sigma_{1p} = \alpha F$, using $\alpha$ from the fits to Eq.~(\ref{eq:saturation}). Although this ignores saturation, it allows us to set an upper bound on $\sigma_{1p}$ for comparison. To convert their excitation photon densities $x_\text{ph}$ to fluence, we note that they define the average density over the excitation volume differently, $x_\text{ph} = (1-1/e)n_\text{ex}$; we also divide by $\zeta$ to correct for their higher temporal resolution. They report their results as changes in optical density, so we compute $\Delta\text{OD} \approx \log_{10}[1-2\Delta E(t_p,0)/E_p] \approx 0.87\alpha\delta^2 Z_0\hbar\omega_p x_\text{ph}/[(1-n)(1-\mathcal{R})(1-1/e)]$ and average the extrapolations for SCOC and YBCO6 to obtain the estimates shown in the table. Considering the uncertainties, the results show reasonable agreement for measurements on both \lco and \nco.
\begin{table}
\caption{Comparison of infrared PIA at 100~meV~\cite{Okamoto2010,Okamoto2011} with the maximum optical density change expected at THz frequencies for the same density. The THz optical density estimates are obtained by extrapolating the present measurements following a procedure described in the text.}
\begin{center}
\begin{ruledtabular}
\begin{tabular}{|c|l|c|c|}
Material & $x_\text{ph}$ & $\Delta\text{OD}$ (100 meV) & max~$\Delta\text{OD}$ (THz)\\\hline
\nco & 0.027 & 0.08 & $0.06\pm 0.01$\\
\lco & 0.055 & 0.05 & $0.11\pm 0.01$
\end{tabular}
\end{ruledtabular}
\end{center}
\label{tab:compare}
\end{table}

A comparison with theory indicates that the mobility is limited by both electron-magnon scattering and electron-phonon scattering. The initial temperature and quantum yield of the photocarriers may also play a role, but we assume these are less important. An exact diagonalization study of the $t$-$J$ model with $J = 0.3t$ shows a low-field limit, $\mu =1.32ea^2/\hbar \approx 3~\text{cm}^2/\text{V\,s}$, that is 4--5 times higher than our results~\cite{Mierzejewski2011}. Including electron-phonon interactions introduces additional scattering that reduces the mobility and improves the agreement. A study of the hole mobility in the $t$-$J$-Holstein model, for example, showed that it decreased by a factor of 2 when the electron-phonon coupling constant increased from $\lambda=0$ to $\lambda=0.3$~\cite{Vidmar2011}. The photoconductivity calculated in the Hubbard-Holstein model shows a similar decrease~\cite{DeFilippis2012}. Both results are within a factor of 2--3 of our peak photoconductivity.

\subsection{Photoconductivity decay}
\label{sec:decay}
The unusually rapid recombination in the insulating cuprates explains a puzzling discrepancy between different measurements of their mobility. Earlier authors measured the dc photoconductance per square $\Delta g_\square$ per incident photon flux $F_\text{ph}$, $\Delta g_\square/F_\text{ph}\approx 10^{-24}~\Omega^{-1}\,\text{cm}^2\,\text{s}$, and estimated the carrier recombination time to be $\tau_r \approx 10$~s from microsecond-resolution photoconductance measurements~\cite{Thio1990}. This gave $\eta\mu = \Delta g_\square/(eF_\text{ph}\tau_r)\approx 10^{-7}~\text{cm}^2/\text{V s}$, where the internal quantum efficiency $\eta$ denotes the fraction of photoexcitations that contribute to the current, equal to the product of the quantum yield $\Phi$ and the free-carrier survival probability over the measurement time. The authors noted that this estimate was orders of magnitude smaller than expected, and suggested that the photocarrier states are more localized than their doped counterparts. Our results show that the photocarrier mobility is comparable to that of doped carriers, but that they recombine over a time scale that is thirteen orders of magnitude shorter than previously assumed. The carriers that remain are less mobile, but they last for a long time, so they dominate in a dc measurement. The contribution during our measurement window, $T\approx 150~\text{ps}$, is a negligible fraction of the total, $\Delta g_\square/F_\text{ph} = (\hbar\omega_p\delta)\int_0^T\sigma(\tau)\,d\tau/F \approx 2\times 10^{-30}~\Omega^{-1}\,\text{cm}^2\,\text{s}$.

The insets to Fig.~\ref{fig:fluence} show that the time dependence of $\sigma_{1p}$ is approximately independent of fluence. As discussed in Sec.~\ref{sec:photocarriers}, this indicates that the decay is dominated by first-order kinetics, since higher-order processes should cause the decay rate to increase with increasing density. In the multimagnon decay model of \textcite{Lenarvcivc2013,Lenarvcivc2014a}, the recombination rate is limited by the decay process and occurs through an exciton state that ensures the kinetics is first-order. In many low-mobility semiconductors, however, the recombination rate is thought to be limited by the time it takes for charge pairs to reach each other~\cite{Pope1999}, and we can derive two pertinent length scales from our results. At high density, the photocarriers become increasingly likely to interact with the oppositely charged carriers of nearby excitations, so the absence of bimolecular recombination would indicate that the photocarriers remain localized as geminate pairs. At low density, the recombination time places an upper bound on the carrier separation that can still produce diffusion-limited recombination.

The mean in-plane separation $\bar{r}_0$ between neighboring photoexcitations at the saturation fluence $F_0$ is less than seven lattice constants in both SCOC and YBCO6, and the encounter radius $r_e$ set by the recombination time is 10--12 lattice constants. Letting $d$ denote the separation between copper oxide planes, and $n_{\text{ex},0} = (1-\mathcal{R})F_0/(\hbar\omega_p\delta)$ the nominal saturation density, $n_{\text{ex},0} \approx 5\times 10^{-3}/\text{Cu}$ in SCOC gives a mean separation $\bar{r}_0 = (1/2)(n_{\text{ex},0}d)^{-1/2} \approx 2.7$~nm, while $n_{\text{ex},0} \approx 6\times 10^{-3}/\text{Cu}$ gives $\bar{r}_0\approx 2.5$~nm in YBCO6. At low density, the diffusion-limited encounter time for an isolated holon-doublon pair separated by $r_e$ is
\begin{equation}
t_e = \int_0^{r_e}\frac{dr}{(\mu_h + \mu_d)E(r)} =  \frac{4\pi\epsilon_r\epsilon_0r_e^3}{3e(\mu_h + \mu_d)},
\end{equation}
which gives $r_e\approx 5$~nm, about twelve lattice constants, for the mobility and initial decay time measured for SCOC, and $r_e\approx 4$~nm, about ten lattice constants, for YBCO6. While neither of these estimates enable us to exclude diffusion-limited recombination in SCOC and YBCO6, they place strong constraints on the initial photocarrier distribution that could produce it.

\subsection{Temperature dependence}
\label{sec:temp}
\begin{figure}
	\centering
	\includegraphics[width=\columnwidth]{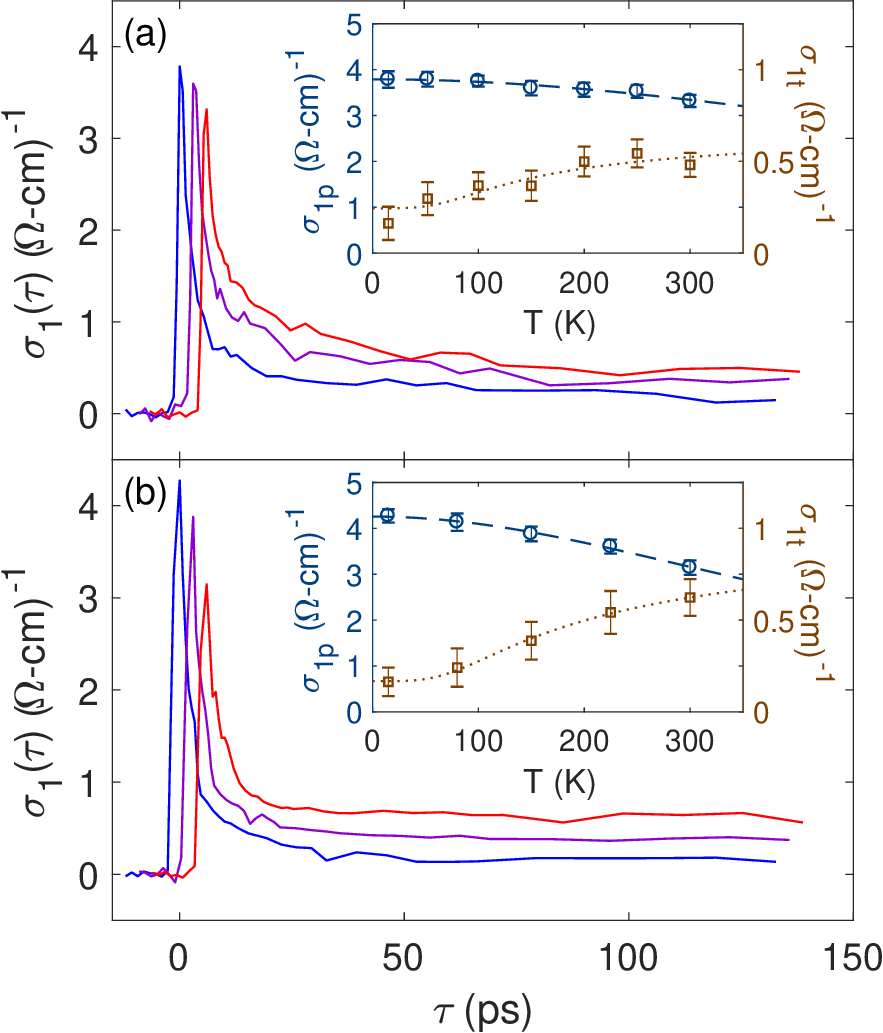}
	\caption{(color online). Time-dependent photoconductivity of SCOC with $F = 240~\mu\text{J/cm}^2$ (a) and YBCO6 with $F = 220~\mu\text{J/cm}^2$ (b), at $T =$~15~K (blue), 150~K (purple), and 300~K (red); for clarity, the curves are shifted along the $\tau$ axis by 0, 3, and 6~ps, respectively. Insets:~Photoconductivity at the peak (dark blue circles, left axis) and over the tail (gold squares, right axis) of the time-dependent photoconductivity curves shown in the main panels, as a function of base temperature $T$. Tail photoconductivity is the average over $100 < \tau < 150~\text{ps}$. Dashed lines show fits of $\sigma_{1p}(T) = 1/(\rho_0 + AT^2)$ to the peak data; dotted lines show fits of $\sigma_{1t} =\sigma_0 \exp(-\varepsilon_A/k_BT_\text{eff})$ to the tail data, using a laser-heated effective temperature $T_\text{eff} = (T^4 + T_\ell^4)^{1/4}$ as described in the text.}
\label{fig:temp}
\end{figure}

The temperature dependence shown in Fig.~\ref{fig:temp} demonstrates that thermal excitations inhibit the THz-PC at early times, but enhance it at longer times.
To understand this, we assume that the recombination processes are independent of temperature, $\sigma_1(\tau,T) = n_c(\tau)e\bar{\mu}(\tau,T)$, and identify $\bar{\mu}(\tau,T)$ with scattering and trapping processes that change with both time and temperature as the carriers evolve toward equilibrium. At fixed time, we assume the carriers form a quasi-equilibrium distribution and treat their temperature dependence as we would in equilibrium. While this is clearly a gross simplification for the state immediately after photoexcitation, we expect it to become more accurate as time progresses. Under these assumptions, the temperature dependence indicates coherent transport at early times, which crosses over to hopping transport at longer times.

We can fit $\sigma_1(\tau,T)$ to standard transport models to make this description more quantitative. To account for the dependence of $T$ on the laser pulse energy, we assume a Debye model for the specific heat, $C\approx \beta T^3$, and relate the absorbed laser energy density $\mathcal{E} \approx F(1-\mathcal{R})/\delta$ to a characteristic temperature $T_\ell = (4\mathcal{E}/\beta)^{1/4}$. A photoexcited region will then reach an effective temperature, $T_\text{eff} = (T^4 + T_\ell^4)^{1/4}$, if all of the laser energy is dissipated thermally. For a Debye temperature $\Theta_D\approx 350$~K~\cite{Eckert1988}, we get $T_\ell\approx 100$~K. Fits with an activation model to the THz-PC at long time delays, $\sigma_{1t} = \langle\sigma_{1}(\tau>100~\text{ps},T)\rangle_\tau = \sigma_0 \exp(-\varepsilon_A/k_BT_\text{eff})$, are shown in the insets to Fig.~\ref{fig:temp}, and yield activation energies $\varepsilon_A = 7\pm 2$~meV for SCOC and $\varepsilon_A = 12\pm 1$~meV for YBCO6, both with $T_\ell \approx 75$~K and $\sigma_0\approx 1~(\Omega\, \text{cm})^{-1}$. These activation energies are remarkably low, much smaller than the characteristic phonon and magnon energies, and suggest hopping in a relatively weak disorder potential.

The insets to Fig.~\ref{fig:temp} also contrast the temperature dependence of $\sigma_{1p}$ with $\sigma_{1t}$, and indicate that the intrinsic photocarrier mobility exhibited at $\tau = 0$ is degraded by equilibrium thermal fluctuations. At these early timescales we can not assume that the laser energy has dissipated into lattice heat, so we simply fit the equilibrium temperature dependence with the phenomenological model $\sigma_{1p}(T) = 1/(\rho_0 + AT^2)$, shown as curves in Fig.~\ref{fig:temp} with best-fit values $A = 390\pm 50~\text{n}\Omega\,\text{cm}/\text{K}^2$ for SCOC and $A = 910\pm 50~\text{n}\Omega\,\text{cm}/\text{K}^2$ for YBCO6. We emphasize here that while this is the same temperature dependence that Fermi liquid theory predicts for electron-electron scattering, other power laws would fit equally well. We use this form because it is a simple analytic expression that captures the temperature dependence.

\section{Conclusions}
\label{sec:conc}
In summary, THz-PC shows photoconductivity in insulating cuprates with a common dependence on fluence, time, and temperature. The peak photoconductivity saturates with fluence, but in the low-density limit we can infer an intrinsic photocarrier mobility of 0.6--0.7~$\text{cm}^2/\text{V}\,\text{s}$. This value is close to the mobility of doped carriers, and is qualitatively consistent with both theoretical predictions and earlier PIA measurements. The photoconductivity decays on multiple picosecond timescales, with an initial decay time of 1--2~ps. Earlier photoconductivity measurements were insensitive at these timescales, so the mobilities estimated from them were much lower.

The decay shows a time dependence that is approximately independent of fluence, indicating that it is governed by first-order kinetic processes. Similar time dependence was observed in the infrared PIA spectral weight, indicating that it is caused primarily by carrier recombination. Our fluence-dependent measurements place limits on the possible recombination pathways, because they include densities for which the mean separation between photoexcitations is only a few lattice constants. The decay time imposes an additional constraint on diffusion-limited recombination mechanisms.

At early times the conductivity decreases with increasing temperature, while at later times it shows the opposite dependence. Assuming that the recombination kinetics is temperature independent and applying a quasi-equilibrium interpretation of the mobility, this indicates that the system crosses over from a coherent regime at early times to a hopping regime at later times. The temperature dependence at selected fixed times can be fit with standard equilibrium transport models, and in the hopping regime this yields activation energies of about 10~meV, indicating a weak disorder potential.

Our work suggests several avenues for future development. While we have shown that our results are broadly consistent with other measurements, numerous gaps remain in the frequencies, temporal resolution, excitation densities, temperatures, and materials that have been studied. Measurements that bridge these gaps would clarify how photocarriers behave in insulating cuprates and establish a broader foundation for more general theoretical and experimental work on interacting electrons out of equilibrium. In particular, the saturation that we have described is observed in many other materials, but the mechanism remains poorly understood. From a purely practical point of view it is often useful to increase the excitation fluence to enhance the signal-to-noise ratio in experiments, so it is important to understand how the saturation limits this approach. It also represents a robust feature of the photoexcitation process that demands a fundamental explanation. Finally, studying the evolution of these effects with chemical doping will strengthen our understanding of the excitations across the phase diagram of the doped antiferromagnetic insulator, allowing new insight into high-temperature superconductivity.

\begin{acknowledgments}
We acknowledge support from NSERC and the CIFAR Program in Quantum Materials, and useful discussions with S.~Kaiser, F.~Hegmann, G.~Sawatzky and J.~Bon\v{c}a.

J.C.P. and A.F.\ contributed equally to this work.
\end{acknowledgments}

\end{document}